\documentclass[aps,nolongbibliography,twocolumn]{revtex4-2}
\usepackage[UKenglish]{babel}
\usepackage{latexsym}

\usepackage[pdftex, colorlinks=true, linkcolor=red, citecolor=blue, urlcolor=magenta, pdfstartview=FitH]{hyperref}
\usepackage{graphicx}	
\usepackage{amssymb}
\usepackage{amsmath}
\usepackage{xcolor}
\usepackage[normalem]{ulem}
\usepackage{xcolor} 

\pdfoutput=1

\usepackage{bm}
\usepackage{verbatim}
\usepackage{amsmath}
\usepackage{amssymb}
\usepackage[utf8]{inputenc}
\usepackage{nicefrac}
\usepackage{braket}
\usepackage{pdfcomment}
\usepackage{pbox}
\usepackage{makecell}
\usepackage{hhline}
\usepackage{multirow}
\usepackage{tabularx}
\usepackage{soul}
\usepackage{float}
\usepackage{microtype}
\usepackage{subcaption}
\usepackage{caption}

\usepackage{subcaption}      
\usepackage{appendix}        
\usepackage{enumitem}        
\usepackage{algorithm}       
\usepackage{algpseudocode}   
\usepackage{booktabs}        
\usepackage{siunitx}         

\newcommand{\authoraff}[2]{#1$^{#2}$}

\begin{document}

\title{Ferroelectric Control of Interlayer Excitons in 3R-MoS$_{2}$ / MoSe$_{2}$ Heterostructures}
%
%
\author{%
\authoraff{Johannes Schwandt-Krause}{1},
\authoraff{Mohammed El Amine Miloudi}{1},
\authoraff{Elena Blundo}{2,3},
\authoraff{Swarup Deb}{1,4,5},
\authoraff{Jan-Niklas Heidkamp}{1},
\authoraff{Kenji Watanabe}{6}, 
\authoraff{Takashi Taniguchi}{7},
\authoraff{Rico Schwartz}{1},
\authoraff{Andreas Stier}{2},
\authoraff{Jonathan J. Finley}{2,3},
\authoraff{Oliver K\"uhn}{1},
\authoraff{Tobias Korn}{1}
}\email{tobias.korn@uni-rostock.de}

\affiliation{$^1$Institute of Physics, University of Rostock, Rostock, Germany}
\affiliation{$^2$Walter Schottky Institute, Technical University of Munich, Munich, Germany}
\affiliation{$^3$Munich Center for Quantum Science and Technology (MCQST), 80799 Munich, Germany}
\affiliation{$^4$Saha Institute of Nuclear Physics, Kolkata, India}
\affiliation{$^5$Homi Bhabha National Institute, Mumbai, India}
\affiliation{$^6$Research Center for Electronic and Optical Materials, NIMS, 1-1 Namiki, Tsukuba 305-0044, Japan}
\affiliation{$^7$Research Center for Materials Nanoarchitectonics, NIMS, 1-1 Namiki, Tsukuba 305-0044, Japan}

\begin{abstract}
    We investigate the interaction between interlayer excitons and ferroelectric domains in hBN-encapsulated 3R-MoS$_2$/MoSe$_2$ heterostructures, combining photoluminescence experiments with density functional theory and many-body Green's function calculations. Low-temperature photoluminescence spectroscopy reveals a strong redshift of the interlayer exciton energy with increasing MoS$_2$ layer thickness, attributed to band renormalization and dielectric effects. We observe local variations in exciton energy that correlate with local ferroelectric domain polarization of the 3R-MoS$_2$ layer, showcasing distinct domain-dependent interlayer exciton transition energies. Gate voltage experiments demonstrate that the interlayer exciton energy can be tuned by electrically induced domain switching. These results highlight the potential for interlayer exciton control by local ferroelectric order and establish a foundation for future ferroelectric optoelectronic devices based on van der Waals heterostructures.
\end{abstract}

\maketitle
\section{Introduction}
The exponential growth of digital information relies on solid-state electronics, where electrons store, transport, and process signals. With increasing device density, energy consumption increases, resulting in efficiency and thermal challenges. Alternative approaches such as spintronics, photonics, and optoelectronics promise routes to lower-power, faster information handling, complementing traditional semiconductor platforms.

Among emerging directions, excitonic devices have gained attention with the advent of two-dimensional (2D) semiconductors such as MoS$_2$, WS$_2$, and their analogues - the transition metal dichalogenides (TMDCs). The strong Coulomb interaction between photo-excited electrons and holes in these atomically thin systems results in tightly bound excitons, enabling strong light-matter interaction~\cite{Chernikov2014,Ugeda2014,Goryca2019} even at room temperature. However, the small exciton Bohr radius leads to short exciton lifetimes and diffusion lengths within monolayers~\cite{Korn_APL10,Marie2016Lifetime,Poellmann2015,KuligChernikov2018diffusion}, posing challenges for device implementation. Novel opportunities to overcome these limits are offered by van der Waals (vdW) heterostructures obtained by stacking TMDC layers~\cite{Geim2013}. There, interlayer excitons (ILX) can form~\cite{Xu_NatComm2015} with electrons and holes in adjacent layers. ILXs offer extended lifetimes and diffusion lengths, making them attractive for optoelectronic applications~\cite{Rivera2018review}. Although exciton transistors have already been demonstrated~\cite{Unuchek_ExcitonTransistor2018}, the concept of storing data encoded in excitonic properties remains largely unexplored.

\begin{figure*}
    \centering
    \includegraphics[width=\textwidth]{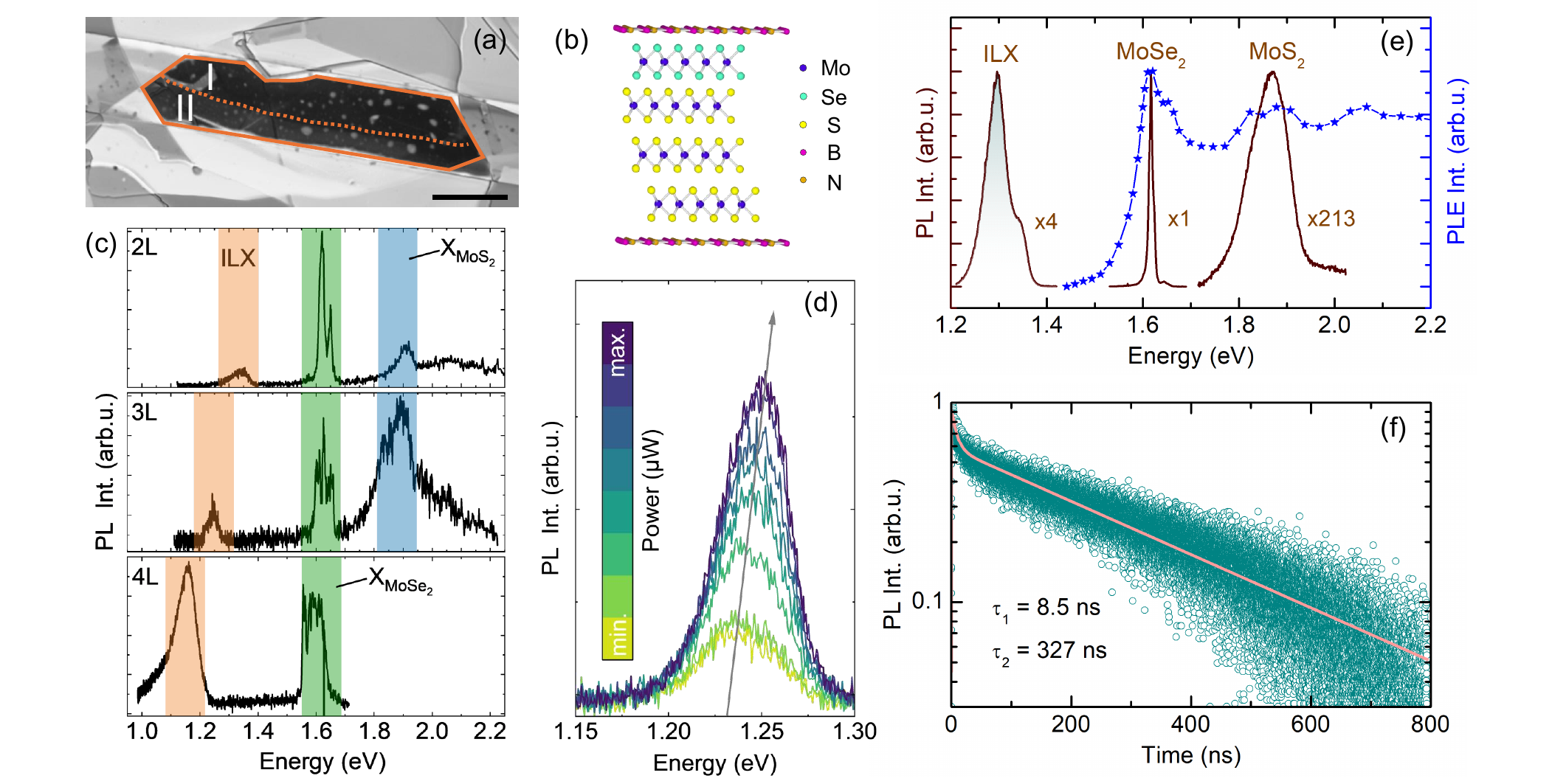}
    \caption{(a) Optical image of an hBN-encapsulated nL-MoS$_{2}$/MoSe$_{2}$ heterostructures on gold contacts. The heterostructure is divided into a region with 3L-MoS$_{2}$ (I) and a region with 4L-MoS$_{2}$ (II), indicated by the orange lines. The black scale bar corresponds to 10 µm. (b) Schematic of the hBN-encapsulated 3L-MoS$_{2}$/MoSe$_{2}$ heterostructure, highlighting the 3R stacking of MoS$_2$. (c) PL signals of different MoS$_{2}$/MoSe$_{2}$ heterostructures with 2L/3L/4L-MoS$_{2}$ from top to bottom. The green and blue areas indicate the intralayer excitons of MoSe$_{2}$ and MoS$_{2}$, respectively. The salmon-colored area shows the ILX signal of the heterostructures. While the 2L and 3L spectra were recorded using a Si CCD, the 4L spectrum was measured using an InGaAs photodiode array. (d) Power dependence of the PL signal in region I of panel (a). The grey arrow indicates a blueshift in the signal with increasing power. (e) Exemplary PL and PLE spectra of an hBN-encapsulated 2L-MoS$_{2}$/MoSe$_{2}$ heterostructure. (f) Time-resolved PL of the ILX feature in the 2L-MoS$_{2}$/MoSe$_{2}$ heterostructure. The pink line is a double exponential fit to the data, yielding the decay times displayed in the figure.}
    \label{Panel1}
\end{figure*}
In this work, we explore excitonic-ferroelectric coupling in 3R-MoS$_2$/MoSe$_2$ heterostructures to show distinct exciton energies dependent on layer thickness and ferroelectric domain polarity.

The emergence of sliding ferroelectricity in TMDCs~\cite{yasuda2021stacking,Li2021sliding,Deb2022Cumulative} has opened up new possibilities for non-volatile device functionalities. Spontaneous ferroelectric polarization occurs in crystals with parallel stacking (3R polytype), unlike the more common anti-parallel (2H) stacking. 
In artificially stacked 3R-layers, ferroelectricity can be engineered via the twist angle~\cite{Sung2020BrokenMirror,Deb2022Cumulative}, however, the fabrication of these structures remains challenging. Alignment uncertainties during the fabrication often limit ferroelectric domain sizes to below 100 nm. Only recently 3R rhombohedrally-stacked TMDCs have been successfully synthesized as single crystals~\cite{Ullah2021selective}. 
While ferroelectric field-effect transistor (FeFET) architectures in 3R-TMDC stacks have shown intriguing transport behavior~\cite{Pablo_FerroFet2020,Ye2018CuInPS,Wei2021CuInPS}, their optoelectronic potential is largely unexplored.

 In this study, we control ILX in 3R-MoS$_2$/MoSe$_2$ using the ferroelectric domain landscape of the 3R-MoS$_2$ layer. The influence of ferroelectricity on \textit{intra}layer excitons in individual TMDCs has been investigated ~\cite{Deb2024}, its effect on ILX in \textit{hetero}structures remains virtually unaddressed. The MoS$_2$/MoSe$_2$ monolayer heterostructure is a well-established platform for hosting ILX~\cite{Plochocka2017MoSeMoS,Ferrari2024,Sokolowski_2023}. However, it remains elusive whether a similar heterostructure incorporating few-layer 3R-MoS$_2$, necessary for the formation of ferroelectric domains, also supports interlayer excitons. We utilize gate voltage-dependent photoluminescence(PL) to probe hBN-encapsulated 3R-MoS$_2$/MoSe$_2$ heterostructures. Our study reveals that 3R-multilayer-MoS$_2$/monolayer-MoSe$_2$ heterostructures can indeed host long-lived and optically bright ILX. Using experimental results and density functional theory (DFT), we identify the influence of ferroelectric domains of the 3R-MoS$_2$ layer on the ILX energy. This sensitivity enables a new approach to exciton control, using the ferroelectric order parameter as a local, reconfigurable, and non-volatile lever to modulate both the spatial distribution and energy levels of excitonic states. These findings make a compelling case for the integration of sliding ferroelectrics into excitonic device architectures, offering a pathway toward optically addressable data storage with electrically reconfigurable control.

\begin{figure*}
    \centering
    \includegraphics[width=1\linewidth]{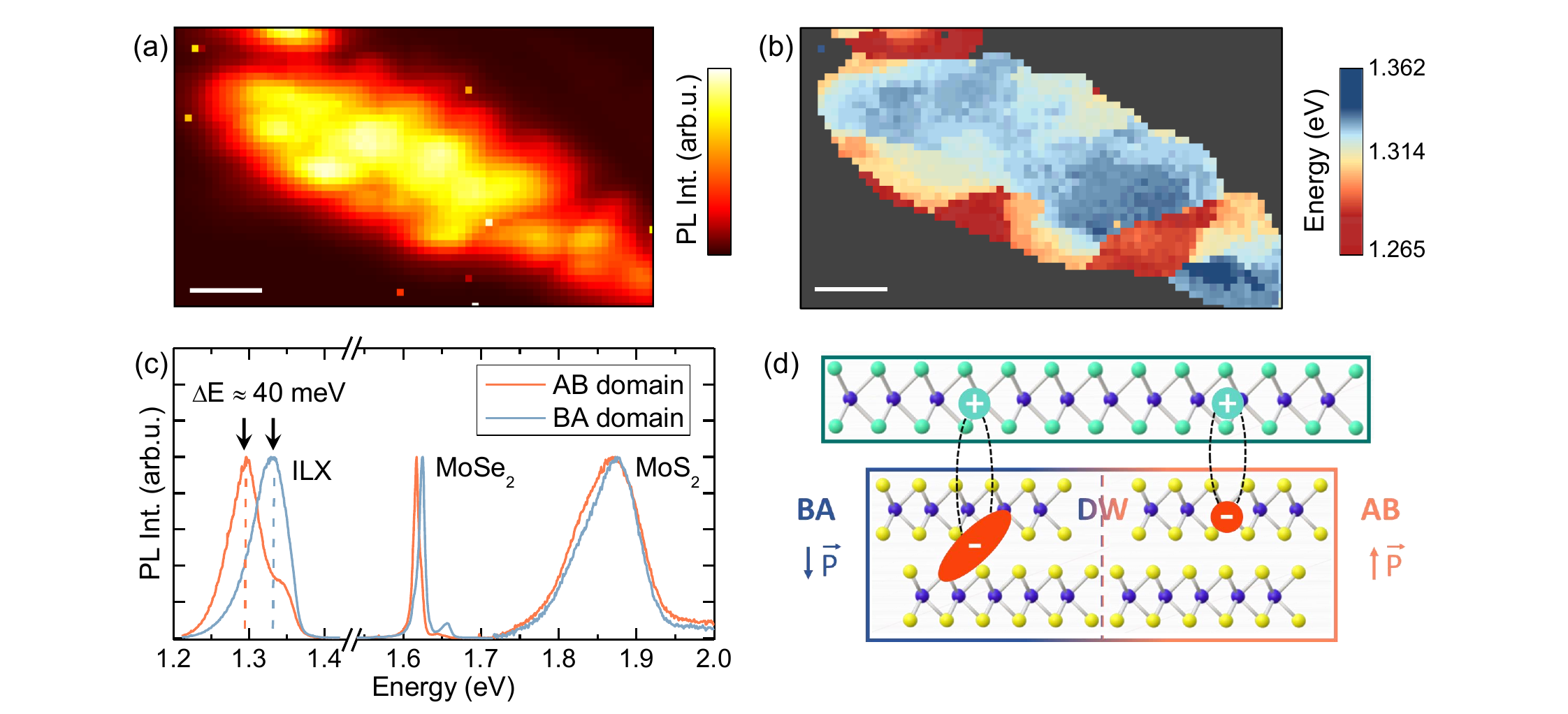}
    \caption{
    (a) ILX PL heatmap from a 2L-MoS$_{2}$/MoSe$_{2}$ heterostructure. (b) Corresponding spatial map of the ILX peak energy in the heterostructure. The white scale bar corresponds to 4 µm. (c) Representative PL spectra from (a), showing the energy difference between AB and BA domains of the underlying 3R-2L-MoS$_{2}$. (d) Illustration of the sample highlighting the ferroelectric effect on the ILX. AB and BA refer to distinct domain configurations with opposite polarizations, and DW denotes the domain wall separating them. A detailed description can be seen in Fig.~\ref{fig:theo1}.}
    \label{Panel2}
\end{figure*}

\section{Results and discussion}
First, we discuss the general structure of our samples. They consist of few-layer flakes exfoliated from bulk 3R-MoS$_2$ crystals and monolayers of MoSe$_2$, stacked at zero twist angle. The layers are encapsulated in top and bottom hBN. In the following, we will refer to these heterostructures as nL-MoS$_{2}$/MoSe$_{2}$, where n is the number of MoS$_2$ layers. 

\subsection{Optical spectroscopy of nL-MoS$_{2}$/MoSe$_{2}$ ($n=2-4$)}

Figure \ref{Panel1}a shows the optical image of one of the hBN-encapsulated 3R-MoS$_{2}$/MoSe$_{2}$ samples, which is divided in a region with a 3L-MoS$_{2}$(I) and a region of 4L-MoS$_{2}$(II). 
A detailed description of the sample preparation can be found in the Methods section. A schematic of the sample (region I) is depicted in Fig.~\ref{Panel1}b, highlighting the 3R stacking of MoS$_2$.  In Fig. \ref{Panel1}c, the PL spectra of three different combinations of nL-MoS$_{2}$/MoSe$_{2}$ heterostructures are shown, with n equal to 2 (top figure), 3 (center) and 4 (bottom).
The colored regions of the plot showcase the different energetic regions of the excitonic features of the heterostructures. 
Noticeably, we observe a distinct ILX feature of the heterostructures close to $1.2~eV$, which exhibits a strong layer-dependent redshift with increasing MoS$_2$ thickness. 
Qualitatively, this can be traced to originate from band renormalization and changes of the excitons dielectric environment in the MoS$_{2}$ layer. As the material switches from a direct to an indirect gap semiconductor with increasing MoS$_2$ thickness, the conduction band minimum and valence band maximum at the $K$-points shift. This change of the MoS$_{2}$ bands causes the band alignment of the heterostructure, and therefore, the transition energy of the ILX, to be altered. Additionally, with increasing $n$, the effective dielectric constant increases, and thus the binding energy of the exciton decreases. This is consistent with previous reports on MoS$_2$/WSe$_2$ heterostructures~\cite{Gao2021Layer-engineerILX}.

Fig. \ref{Panel1}d shows power-dependent ILX emission for the 3L-MoS$_{2}$/MoSe$_{2}$ heterostructure. A clear blueshift of the ILX emission peak with increasing laser power is observed, which can be attributed to the competing effects of dipole–dipole repulsion~\cite{Xu_NatComm2015,Wurstbauer17,Nagler17} and attractive screening-induced self-energy corrections at higher exciton densities~\cite{Chernikov24}—a hallmark of the interlayer nature of this PL feature.

In Fig.~\ref{Panel1}e and f, PL excitation (PLE) and time-resolved PL measurements of the ILX in a 2L-MoS$_{2}$/MoSe$_{2}$ heterostructure are shown, respectively. The PLE measurements clearly show resonant behavior with the monolayer MoSe$_{2}$ and MoS$_{2}$ intralayer excitons, demonstrating good electronic coupling and charge transfer between the constituent layers of the heterostructure\cite{Xu_NatComm2015,Nagler17,Blundo2024, Hanbicki2018,Soubelet2024PolaronsILX,Blundo2025InSeILX}. The time-resolved PL for this ILX reveals a biexponential decay with time constants of $\tau_1=8.5 ~ns$ and $\tau_2=327~ns$. Previous work on MoS$_{2}$/MoSe$_{2}$ heterobilayers also reported biexponential decay transients with decay times in the range of 0.1 to 3 $\mu$s~\cite{Ferrari2024}. Given that ILX lifetimes in TMDC-based heterostructures are known to be highly dependable on sample specifics such as twist-angle and excitation power~\cite{Blundo2024,Choi2021_TwistAngleILX}, our ILX lifetimes agree reasonably well with previous results, and furthermore underscore the interlayer character of the detected signal, as they exceed intralayer exciton lifetimes by several orders of magnitude~\cite{Korn_APL10,Robert2016_Phys_Rev}. 

We continue with investigating the domain structure of the heterostructures utilizing spatially resolved optical spectroscopy. 
Fig.~\ref{Panel2}a and \ref{Panel2}b show false color maps of the ILX PL intensity and ILX transition energy, respectively, of a 2L-MoS$_{2}$/MoSe$_{2}$ heterostructure. 
While there are no discernible systematic ILX intensity variations across the sample, clear regions of distinct ILX transition energy are observed.
The pronounced transition energy shift of about 40 meV between different domains is exemplified by the spectra in Fig. \ref{Panel2}c. This position dependence can not be explained by a change from R-type to H-type interlayer alignment between the MoS$_2$ and MoSe$_2$, as the constituent flakes are single crystals (confirmed by Second Harmonic Generation measurements, SI Fig. S6). The observed energy shift in the PL spectra appears to be closely tied to ferroelectric domains in the 2L-MoS$_{2}$. Fig. \ref{Panel2}d schematically depicts the effect: for rhombohedrally grown 2L-MoS$_2$, there are two stable configurations with opposite ferroelectric polarization, called AB (polarization pointing down) and BA (polarization pointing up). 
The ferroelectric polarization leads to an energetic shift of the MoS$_2$ conduction band in the layer adjacent to the MoSe$_2$. Furthermore, the FE polarization influences the localization of the electron bound in an ILX. Consequently, the total energy of the ILX changes, reflected in the observed ILX transition energy. Our experiment, therefore, establishes a foundation for ferroelectric field-effect devices (FeFED) in which information is stored via ferroelectric polarization states, transcribed directly into the excitonic energy.
We note that due to the difference in lattice constants between MoS$_2$ and MoSe$_2$, the interface between the two materials does not yield extended regions with well-defined interlayer registry: even for parallel alignment of the layers, a moiré pattern with a wavelength of about 8~nm develops. Thus, there are no extended domains with defined ferroelectric polarization at this interface but only within the 3R-MoS$_2$ layer.

\subsection{DFT/BSE Results for 2L-MoS$_{2}$/MoSe$_{2}$}

\begin{figure*}
    \centering
    \includegraphics[width=0.9\linewidth]{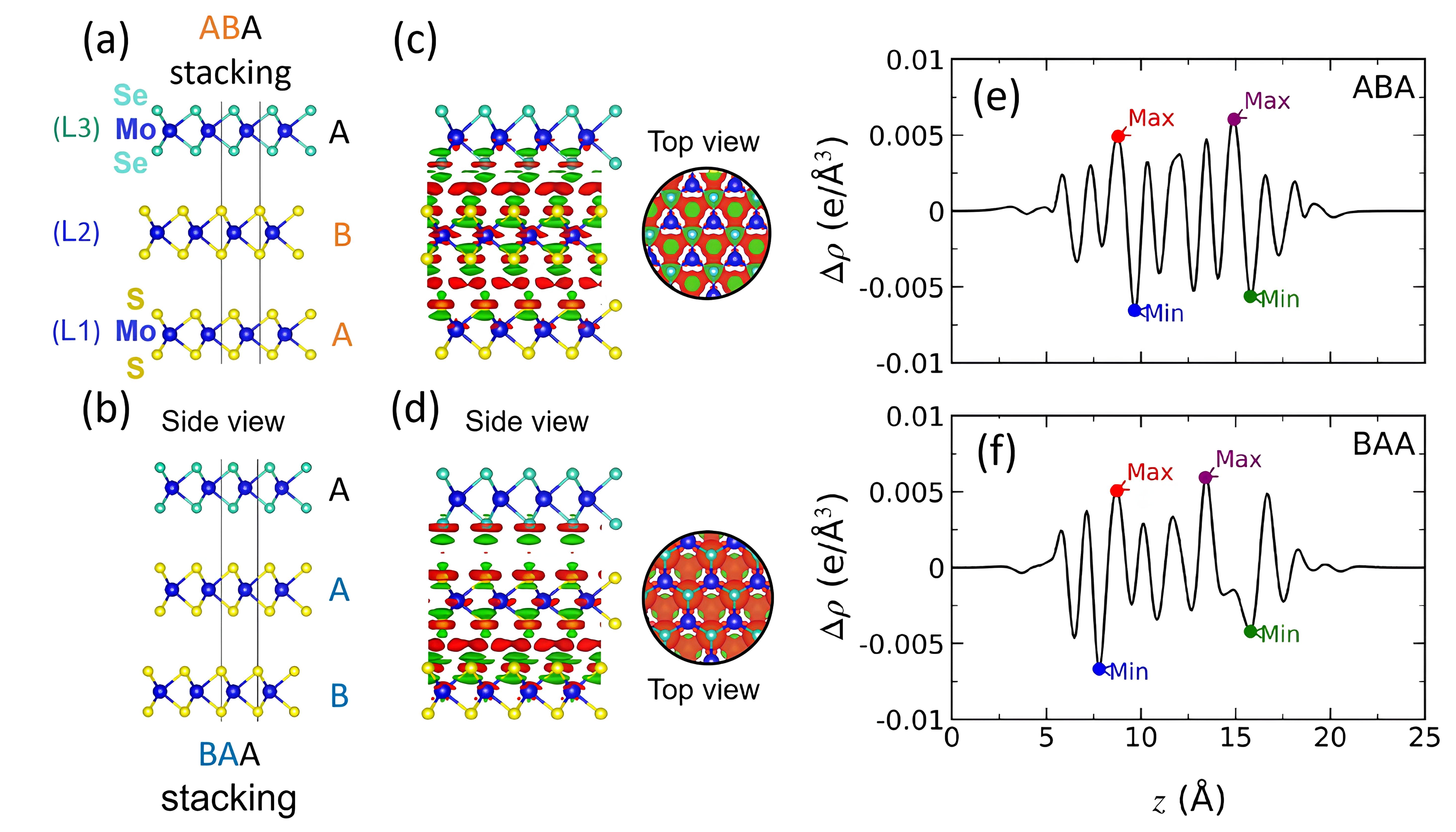}
    \caption{(a,b) Side views of the ABA and BAA stackings, respectively.
   The two parallel black lines denote the unit cell. (c,d) Side and top views of the charge density difference isosurfaces for ABA and BAA stackings, respectively, highlighting interfacial dipoles and polarization reversal; red and green isosurfaces represent charge depletion (holes) and accumulation (electrons), respectively, at an isovalue of 0.001 e/\AA$^{3}$. (e,f) In-plane averaged charge density difference profiles along the $z$-axis for ABA and BAA stackings, respectively, with the 2L-MoS$_2$ (AB)  in ABA exhibiting a Max$\to$Min sequence (Max $\approx$ +0.005 e/\AA$^{3}$, Min $\approx$ -0.0065 e/\AA$^{3}$) and BA is showing an inverted  ordering. }
    \label{fig:theo1}
\end{figure*}

To support the experimentally observed dependence of the ILX transition energy on the ferroelectric domains, we performed  DFT calculations for the exemplary case of 2L-MoS$_2$/MoSe$_2$. Specifically, we examined seven high-symmetry stacking configurations, i.e. AAA, AAB, ABB, BBA, BAA, BAB, and ABA, generated by laterally shifting the MoSe$_2$ or MoS$_2$ layers relative to one another, starting from the perfectly aligned AAA configuration. The corresponding atomic registries are shown in SI Fig.~S1a-g. In the following, we will focus on the energetically favorable ABA and BAA configurations shown in Fig.~\ref{fig:theo1}a-b; the additional results for the other stacking configurations are shown in the SI. The fact that the relative stacking of the MoSe$_2$ layer is not well-defined due to the lattice mismatch will be addressed subsequently below.

Figure~\ref{fig:theo1}c and \ref{fig:theo1}d show isosurfaces of the (ground state) charge density difference, revealing the interfacial dipoles characterized by regions of charge depletion and accumulation. The comparison of the two panels confirms polarization reversal, i.e. the ABA stacking exhibits a 2L-MoS$_2$ interlayer dipole oriented downwards, whereas BAA shows the opposite orientation.  A more quantitative picture is provided by the in-plane-averaged charge density difference profiles, $\Delta \rho(z)$, given in  Fig.~\ref{fig:theo1}e-f. Here, the ABA stacking for the MoS$_2$ bilayer shows a Max$\to$Min sequence (from $+0.0050$ to $-0.0065~e/\text{\AA}^3$), whereas BAA exhibits the reverse pattern.  Overall, this behavior is consistent with the spontaneous polarization switching characteristic of bilayer MoS$_{2}$ ferroelectricity. To examine the interplay between stacking and ferroelectricity, we also evaluated the spontaneous polarization ($\vec{P}$) in the constituent MoS$_2$ bilayers and the full heterostructure. The AB-stacked bilayer exhibits an out-of-plane polarization of $7.94 \times 10^{-10}$~C/m ($ \vec{P}_\downarrow$), which reverses in the BA stacking, while the AA configuration is non-polar. At the MoS$_2$/MoSe$_2$ interface, the polarization is markedly reduced by almost an order of magnitude to $0.98 \times 10^{-10}$~C/m, consistent with suppressed interlayer hybridization due to chemical asymmetry and lattice mismatch. 
These results suggest that the net out-of-plane polarization in the heterostructure is primarily inherited from the MoS$_2$ bilayer, with minimal contribution from the interface to MoSe$_2$.

\begin{figure*}
    \centering
    \includegraphics[width=0.8\textwidth]{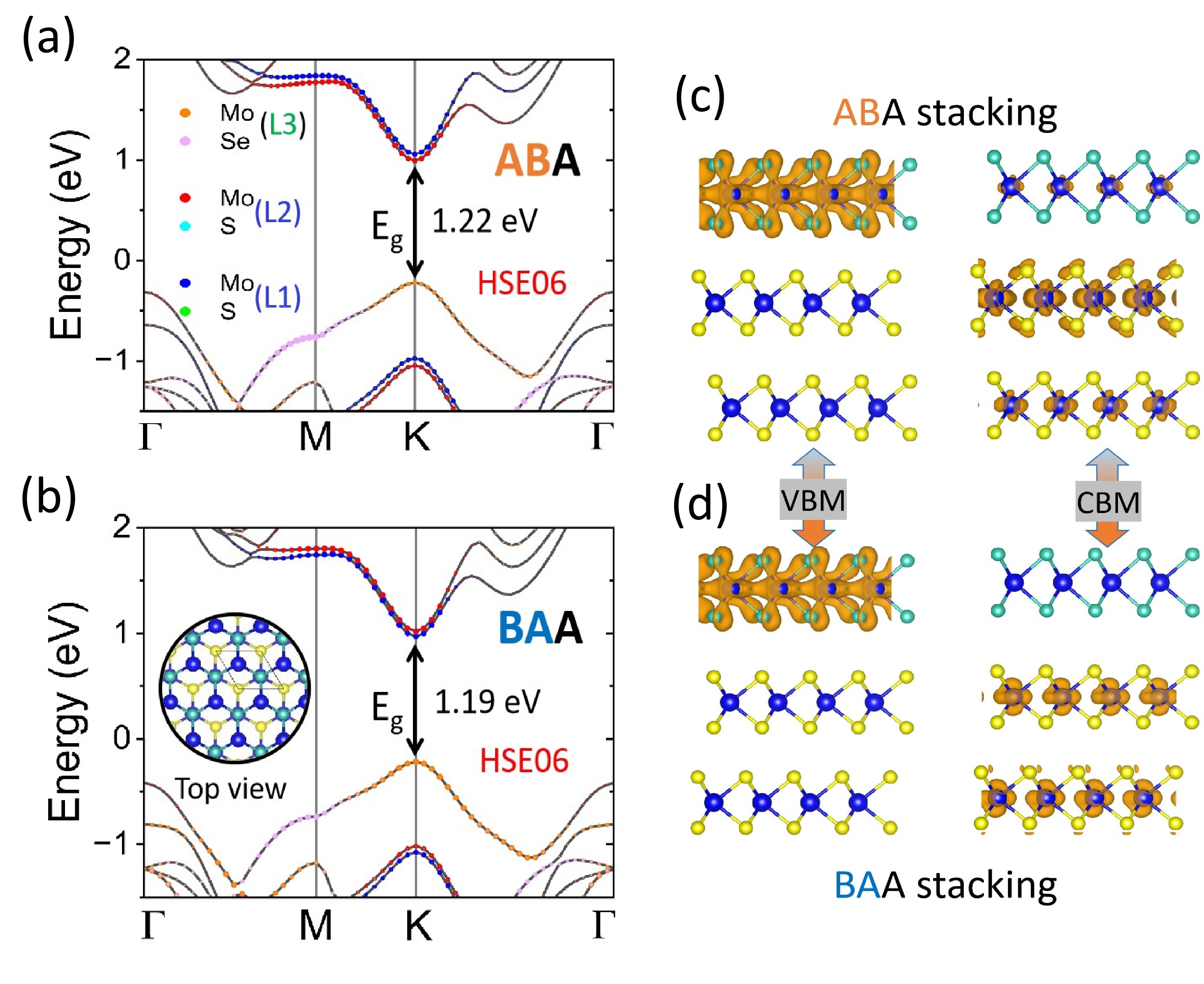}
    \caption{ (a, b) Band structures of 2L-MoS$_{2}$/MoSe$_{2}$ for the ABA and BAA stacking configurations. A top view of the stacking geometry, which is identical for both configurations, is included as an inset in panel (b) for reference. (c,d) Exciton charge density distributions of the CBM and VBM states for ABA and BAA stackings as indicated.}
    \label{fig:theo2}
\end{figure*}

To further interpret the domain-dependent PL shifts observed in Fig.~\ref{Panel2}, we analyzed the electronic structure of the selected stacking motifs representative of the experimentally relevant ABA and BAA domains. The calculated band structures are shown in Fig.~\ref{fig:theo2}a,b.  Both configurations exhibit type-II band alignment, with the valence band maximum (VBM) predominantly localized in MoSe$_2$ and the conduction band minimum (CBM) in MoS$_2$, as detailed in SI Fig.~S2.  Hybrid functional (HSE06) calculations yield direct band gaps at the $K$ point with E$_g$=1.69~eV for ABA and 1.72~eV for BAA.

The heterostructure band gap (E$_g$) can be disentangled into the band gaps of the different layers (L1 to L3, cf. Fig.~\ref{fig:theo1}a). In the ABA/BAA stacking we find E$_\mathrm{g1}=2.02/2.05$~eV, E$_\mathrm{g2}=2.04/2.04$~eV, and E$_\mathrm{g3}=2.33/2.35$~eV. Thus, in the ABA configuration, the total band gap is formed by the CBM of the  MoS$_2$ layer (L2) and the VBM of the  MoSe$_2$ layer (L3). In contrast, the BAA configuration features a band gap formed between the CBM of the MoS$_2$ layer (L1) and the VBM of L3. This change of the CBM from L2 and L1 can be attributed to the effect of reversing the polarization and thus the local electrostatic potential.  Layer-resolved CBM analysis reveals energetic asymmetries. ABA's first MoS$_{2}$ layer CBM is at 1.058~eV while the second layer's CBM is at 0.995~eV, resulting in a 63~meV splitting. Conversely, BAA reverses this offset, with the first layer at 0.971~eV and second at 1.020~eV, corresponding to a -49~meV offset.

PL energies have been obtained according to the lowest eigenvalue of the Bethe-Salpeter equation. Note that these transitions have rather low oscillator strengths and thus can be attributed to ILX. Higher-energy excitations with significantly larger oscillator strengths, corresponding to monolayer (intralayer) excitons appear near 1.9~eV with comparable intensities (cf. SI Tab. S1).
In Table~\ref{tab:PL_essential} we compare results for the  2L-MoS$_{2}$/MoSe$_{2}$ with 1L-MoS$_{2}$/MoSe$_{2}$ and bilayer MoS$_{2}$. Throughout, we observe stacking-dependent shifts of the PL energy. The smallest shift is found for the MoS$_{2}$ bilayer case (12 meV) whereas for 1L- and 2L-MoS$_{2}$/MoSe$_{2}$ the ILX energy shifts are 107 and 69~meV, respectively. Beyond band structure renormalization, our results underscore the critical influence of dielectric screening and stacking-dependent electronic structure. As the number of MoS$_2$ layers increases, the effective dielectric constant increases, leading to lower exciton binding energies and shifted band edges. These effects collectively contribute to the reduced ILX transition energies and confirm that both structural and electronic factors govern the excitonic properties of nL-MoS$_2$/MoSe$_2$ heterostructures.

\begin{table}[ht]
\centering
\caption{
PL energies of the lowest-energy momentum-direct 
ILX transitions in bilayer MoS$_2$, 1L- and 2L-MoS$_2$/MoSe$_2$ heterostructures, obtained from BSE@G$_0$W$_0$ calculations. These correspond to the first optical excitation and exhibit very low oscillator strengths (typically $<0.1$), consistent with the ILX character.
}
\label{tab:PL_essential}
\begin{tabular}{l|c|c}
\hline\hline
\textbf{System} & \textbf{Stacking} & \textbf{ILX PL Energy (eV)} \\
\hline
Bilayer MoS$_2$ & AB & 2.085 \\
                & BA & 2.073 \\
\hline
MoS$_2$/MoSe$_2$ & AB & 1.445 \\
                 & BA & 1.338 \\
\hline
2L-MoS$_2$/MoSe$_2$ & BAA & 1.356 \\
                    & ABA & 1.287 \\
\hline\hline
\end{tabular}
\end{table}

These differences between BAA and ABA 2L-MoS$_2$/MoSe$_2$ PL energies are reflected in the real space exciton densities. In Fig.~\ref{fig:theo2}c,d, the densities are shown for the CBM and VBM projections. As expected for the ILX  the hole density is concentrated in the MoSe$_2$ layer, whereas the electron density spreads over the two MoS$_2$ layers.  Notably the distribution of electron density is found to depend on the stacking.
From Fig.~\ref{Panel2}c, the PL energies of the ILX are obtained as 1.30 and 1.34~eV for AB(A) and BA(A) stacking, respectively.  The calculated values of 1.287 and 1.356~eV are in very good agreement with these values. 
The sign of the splitting is properly reproduced whereas its absolute value is overestimated, i.e. 40~meV vs 69~meV, but still reasonable given the fact that, for instance, the G$_0$W$_0$ approximation is used. One might argue that the experimental results are influenced by lattice mismatch between MoS$_2$ and MoSe$_2$ and domain disorder that leads to spatial averaging over different local stacking configurations with respect to MoSe$_2$. To account for the variable stacking configuration between MoS$_2$ and MoSe$_2$  we calculated the average PL energy of the ABA and  ABB stackings (1.339\,eV) and of the  BAA and BAB stackings (1.376~eV). The resulting splitting is 37~meV, which even improves the comparison with experiment, thus pointing to the significance of spatial variations of the local atomic registry. 

\begin{figure*}
    \centering
    \includegraphics[width=\textwidth]{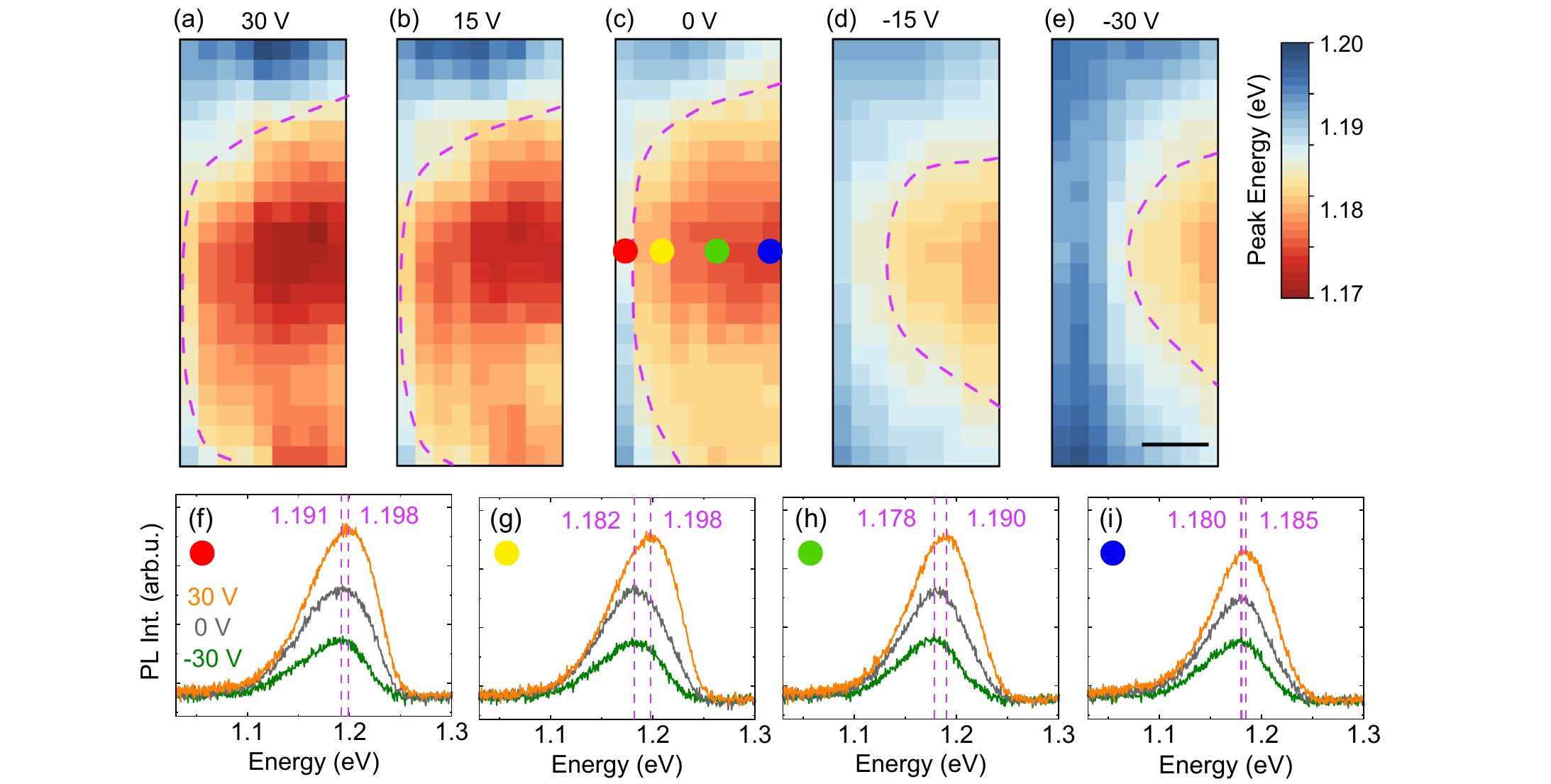}
    \caption{(a-e) Gated PL false color maps of the ILX peak energy in a 3L-MoS$_{2}$/MoSe$_{2}$ heterostructure. The purple dashed line is a guideline to highlight the boundary edge between two domains in the different false color maps. The white scale bar corresponds to 2 µm. (f-i) PL spectra corresponding to the positions highlighted by the coloured dots in panel (c). The pink dotted lines show the spectral centroid range of the spectra, which is also written next to the lines in eV. While the change in the spectral centroid is smaller on the extreme positions (red and blue dots) of the false color map ($\approx$ 5-7 meV), the spectra in the middle of the false color maps (yellow and green dots) show more substantial change ($\approx$ 12-16 meV).}
    \label{Panel3}
\end{figure*}

\subsection{External Field Control of the ILX}
Building on the spectroscopic and DFT evidence for ILX sensitivity to the ferroelectric domain landscape, we now investigate how an externally applied electric field in a FeFED geometry allows, as a proof-of-concept, to achieve non-volatile, electrically reconfigurable exciton control. For this purpose, a contacted 3L-MoS$_2$/MoSe$_2$ heterostructure was fabricated with a bottom gate and spectroscopically probed. Gating the sample affects the charge carrier concentration by doping the sample slightly in the process. More significantly, as demonstrated in our earlier work~\cite{Deb2024}, varying the bottom-gate voltage can alter the ferroelectric polarization through domain-wall motion, thereby reshaping the local polarization state and the overall domain landscape. This mechanism differs subtly from the concept of polarization switching via microscopic in-plane sliding of individual layers relative to one another~\cite{sui2024atomic}. In Fig.~\ref{Panel3}a-e, the PL peak energy of the ILX is shown for different gate voltages. The purple dashed line indicates the domain boundary between different ferroelectric polarizations. The change of the local ferroelectric polarization is evident from the shrinkage of the blue area (expanding red area) with increasing gate voltage. Increasing the gate voltage causes the domain wall to slide so as to maximize the area of the energetically favorable stacking. To make the observation explicit, exemplary spectra from Fig.~\ref{Panel3}a-e (marked in Fig.~\ref{Panel3}c) are shown in Fig.~\ref{Panel3}f-i. At the extreme positions (red and blue circles) of the scans, the energy of the ILX is changing only slightly with gate voltage. In contrast, at the central positions (green and yellow circles), the energy change of the ILX is larger compared to the extremes, indicative of the fact that the domain wall swept through the probe spot as the gate voltage was changed. Above, we discussed that the ILX in a 2L-MoS$_2$/MoSe$_2$ heterostructure experiences a change of peak energy of about 40~meV (see Fig.~\ref{Panel2}b,c) between different domains. The change in energy in the 3L-MoS$_2$/MoSe$_2$ heterostructure (Fig.~\ref{Panel3}a-e) is comparatively smaller. A quantitative reasoning of this contrasting behavior would require \textit{ab initio} calculations on larger layer counts for the 3R-MoS$_2$ part of the heterostructure, which is computationally too expensive for the present study. Moreover, detailed device-level modeling and fabrication/characterization of the FeFED are outside the scope of this work. Consequently, we limit ourselves here to a conceptual interpretation of the gating results. Our theoretical calculations in Fig.~\ref{fig:theo2}c-d show that the electron component of the ILX density is largely delocalized across the entire MoS$_2$ stack, with ferroelectric polarization introducing only a partial asymmetry. It is this stacking-induced change in asymmetry that results in the domain-specific ILX spectra. In the trilayer case, the change in asymmetry is expected to be smaller, because the electron density spreads across more layers, while the domain wall crossing alters the stacking order (therefore, the ferroelectric potential) across a single interface. As a result, domain-to-domain shifts in the ILX energy are reduced. Additional electrostatic screening from gate-doped carriers in the device can further wash out the energy contrast between domains. 

Our results, therefore, reveal an important and counterintuitive design consideration for FeFED optoelectronics: although the total ferroelectric polarization increases with additional 3R-MoS$_2$ layers, the ILX energy becomes less sensitive to local ferroelectric domains. This finding should be borne in mind when engineering layer stacks for electrically reconfigurable excitonic devices.

\section{Conclusion}
In summary, we have fabricated 3R-nL-MoS$_{2}$/MoSe$_{2}$ heterostructures 
that host optically bright ILX, whose energies redshift with increasing MoS$_2$ layer count. These ILX are affected by the intrinsic ferroelectric domain landscape of the 3R-MoS$_{2}$. Using DFT/BSE calculations, we validated the band alignment of the different heterostructures, the effect of ferroelectricity on the band structure as well as the localization of the bound electron in the ILX state. Applying gate voltages allowed us to alter the size of local ferroelectric domains and thus change the energy landscape of the ILX. Therefore, our study paves the way for novel non-volatile reconfigurable optoelectronic devices based on interfacial ferroelectricity.

\section{Methods}
\subsection{Heterostructure fabrication}
Thin Au/Ti (16/3~nm) electrodes were realized by a maskless photolithography system (Smartprint) and  metal evaporation onto silicon substrates covered by 300~nm of SiO$_{2}$. The heterostructure was then fabricated using a polycarbonate (PC)/polydimethylsiloxane (PDMS) - assisted heated transfer of few-layer of hBN and 3R-MoS$_{2}$ and monolayers of MoSe$_{2}$ (HQGraphene). Flakes were first exfoliated and then optically selected (by comparing their contrast to flakes of known thickness) on SiO$_{2}$ (hBN) or on PDMS(MoS$_{2}$/MoSe$_{2}$). The few-layers MoS$_{2}$ and monolayers of MoSe$_{2}$ were then aligned to each other and stamped onto one PDMS film. A PC/PDMS stack was then prepared and used as transfer agent with the van der Waals pick up method ~\cite{zomer2014fast,Pizzocchero2016}. The heterostack was carefully aligned with the already picked up hBN on a PC/PDMS glass slide and lastly dropped on the electrodes of the silicon substrate. The completed stack was then washed in chloroform for 1~h and subsequentially thermally annealed in forming gas at 180~°C for 3~hours. 

\subsection{Continuous-wave $\mu$-PL mapping measurements}
For the PL mapping measurements two different setups were used: 

Setup 1 (University of Rostock): The samples were excited with a 2.33~eV continuous-wave diode laser focused to a spot size of about 1.5~$\mu$m using an 50x microscope objective. The samples were mounted in a helium-flow cryostat (CryoVac Konti Micro) and cooled to a temperature of 4~K. The PL light emitted by the sample is collected using the same objective, filtered by long-pass filters and analyzed with a combination of a spectrometer (Princeton Acton SP2300), a charged-coupled device (CCD) and an InGaAs photodiode array (Princeton PyLoN-IR). The InGaAs photodiode array was cooled by liquid nitrogen to 200~K 
during all experiments. To obtain PL maps of the samples, the cryostat, with the samples inside, was moved in relation to the fixed laser spot through a computer-controlled xy stage. For the gating experiments, the samples were contacted using a gold-wire bonder on a custom copper PCB, which was connected to pins inside the cryostat. By using a Fischer connector and a break-out-box the connections were transmitted to a Keithley 2450 Sourcemeter, which was controlled remotely by a computer.

Setup 2 (TU Munich): For $\mu$-PL maps a single-frequency 633~nm laser diode was used for the excitation. Low laser powers of about 2~$\mu$W focused through a microscope objective with high numerical aperture (NA = 0.82) were employed. xy scanners (by Attocube) were employed to move the sample during the map.
The luminescence signal was collected in a backscattering configuration and was spectrally dispersed by a 0.5~m focal length monochromator (HRS500 by Princeton Instruments) equipped with a 300 grooves/mm grating (with blaze at 1~$\mu$m). The signal was then detected by an etalon-free electrically-cooled Si CCD camera (Blaze 100HRX by Princeton Instruments). The laser light was filtered out by a long-pass filter at 650~nm. The sample temperature was about 7 K during the measurement. 
\subsection{Time-resolved $\mu$-PL measurements}
For time-resolved $\mu$-PL measurements, the sample was excited with a ps supercontinuum laser (NKT Photonics) tuned at about 530~nm, with a full width at half maximum of about 8~nm and 50~ps pulses at 1.2~MHz repetition rate. A high NA (0.75) objective was employed to focus the laser beam (with 0.5 $\mu$W laser power) and collect the signal in a backscattering configuration. The desired spectral region corresponding to the interlayer exciton emission was selected through longpass and shortpass filters, and the signal was collected through an avalanche photodetector from MPD with temporal resolution of 30~ps. The sample temperature was about 6~K during the measurement.
\noindent
\subsection{$\mu$-PL excitation measurements}
For $\mu$-PL excitation ($\mu$-PLE), the same ps supercontinuum laser (with 77.8~MHz repetition rate) and objective used for time-resolved $\mu$-PL were employed. The laser wavelength was changed by an acousto-optic tunable filter, resulting in a bandwidth of about 8~nm. Shortpass and longpass filters were used when appropriate to remove spurious signals from the laser. A laser power of 40~$\mu$W  was employed. The luminescence signal was collected in a backscattering configuration, through a 0.2~m focal length monochromator (by Princeton Instruments) equipped with a 300 grooves/mm grating (with blaze at 1~$\mu$m), and an etalon-free N$_{2}$-cooled Si CCD camera (100BRX by Princeton Instruments). The laser light was filtered out by long-pass filters. The sample temperature was about 6~K during the measurement.

\subsection{Second harmonic generation measurements}
The measurements were performed by using a tunable pulsed Ti:Sapphire laser at 925~nm, with a pulse width of $<100$~fs and a repetition rate of 80~MHz. The sample was excited and measured with a high NA objective (0.9), and the frequency-doubled light was collected in a back\-scattering configuration. A linear polarizer was used to select the experimental configuration with co-polarized laser and SHG signal. A half waveplate mounted on motorized rotation stage was used  for polarization-resolved measurements. The signal was collected through a 0.2~m focal length monochromator (by Princeton Instruments) equipped with a 300 grooves/mm grating (with blaze at 1~$\mu$m), and a etalon-free N$_2$-cooled Si CCD camera (100BRX by Princeton Instruments). The sample was kept at room temperature during the measurement.

\subsection{Computational Details}

Density functional theory (DFT) calculations with the projector augmented-wave (PAW) method \cite{blochl94_17953a, kresse99_1758} were performed using the Vienna Ab initio Simulation Package (VASP) \cite{kresse96_11169a, kresse96_15a}.  Exchange–correlation effects were treated using the Perdew–Burke–Ernzerhof (PBE) functional within the generalized gradient approximation (GGA), including van der Waals interactions via the D3 dispersion correction by Grimme (PBE+D3) \cite{perdew96_3865a, grimme10_154104}. 
PBE was used for structural relaxations and ground-state electronic structure calculations. To improve the accuracy of band gaps, hybrid functional calculations were also performed using the Heyd–Scuseria–Ernzerhof functional (HSE06). 
A plane-wave energy cutoff of 450 eV was used for all calculations. The Brillouin zone was sampled using a $\Gamma$-centered Monkhorst–Pack mesh of $12 \times 12 \times 1$ \textbf{k}-points. A vacuum spacing greater than 20 \AA\ was included along the out-of-plane direction to prevent spurious interlayer interactions.

To construct the heterostructure models, we first optimized the in-plane lattice constants of the individual monolayers, obtaining 3.166~\AA\ for MoS$_2$ and 3.295~\AA\ for MoSe$_2$, corresponding to a lattice mismatch of approximately 4.1\%. To minimize strain and ensure periodic boundary conditions, we imposed a common in-plane lattice constant equal to the arithmetic mean of the two values, resulting in $a = 3.231$~\AA. This leads to a symmetric biaxial strain of about +2.1\% in MoS$_2$ and $-1.9$\% in MoSe$_2$. All atomic positions were relaxed while keeping the in-plane lattice fixed.
To assess alternative modeling strategies, we also tested several commensurate supercell combinations, including 4$\times$4 MoS$_2$ / 3$\times$3 MoSe$_2$, 5$\times$5 / 4$\times$4, 7$\times$7 / 6$\times$6, and 9$\times$9 / 8$\times$8 configurations. In all cases, the residual mismatch between the two layers remained greater than 7\%, even for relatively large supercells, rendering them computationally expensive and unsuitable for systematic exploration of stacking-dependent effects. Moreover, the introduction of large supercells would complicate the analysis by folding the Brillouin zone and obscuring direct comparison across configurations. Therefore, we adopted the primitive unit cell with an averaged lattice parameter as a computationally efficient and physically justified approximation.

Ferroelectric polarization was computed using the Berry phase formalism \cite{king-smith93_1651, resta94_899, xiao10_1959}, employing a dense $16 \times 16 \times 1$ \textbf{k}-point mesh. 

Many-body perturbation theory was employed to calculate quasiparticle and optical excitation properties. Quasiparticle energies were obtained using single-shot G$_0$W$_0$ calculations based on PBE.
Optical absorption spectra were calculated by solving the Bethe–Salpeter equation (BSE), which accounts for correlated electron–hole excitations \cite{hanke1980_prb, strinati1988_rnc}. For the GW and BSE calculations, we employed PAW pseudopotentials specifically optimized for many-body perturbation theory (GW PAW potentials), which include additional semicore states in the valence configuration.
 All G$_0$W$_0$ and BSE calculations were performed using a $\Gamma$-centered \(12 \times 12 \times 1\) \textbf{k}-point grid. The number of unoccupied bands  was set to 32 for bilayer MoS$_2$ and MoS$_2$/MoSe$_2$, and increased to 64 for the 2L-MoS$_2$/MoSe$_2$ systems. More details on the computational setup can be found in Section S1 of the SI.

\section{Acknowledgements}
The authors gratefully acknowledge technical assistance and fruitful discussions with J. Schröer. E.B. gratefully acknowledges the German Science Foundation (DFG) for financial support via the Cluster of Excellence Munich Center for Quantum Science and Technology (MCQST, EXC2111) via the distinguished postdoc program. J.J.F. gratefully acknowledges the German Science Foundation (DFG) for financial support via the Clusters of Excellence Munich Center for Quantum Science and Technology (MCQST, EXC2111) and e-conversion (EXC 2089), as well as the Priority Programme SPP-2244 (FI947/7-2) and the large instrumentation fund via INST 95/1642-1. T.K. acknowledges financial support by the DFG \emph{via} the following
grants: SFB1477 (project No. 441234705), KO3612/5-1 (project No.
398761088) and KO3612/8-1 (project No. 549364913).
K.W. and T.T. acknowledge support
from the JSPS KAKENHI (grant numbers 21H05233 and
23H02052) and World Premier International Research
Center Initiative (WPI), MEXT, Japan.

\clearpage
\onecolumngrid
\newpage
\renewcommand{\thefigure}{S\arabic{figure}}
\renewcommand{\thetable}{S\arabic{table}}
\setcounter{figure}{0}
\setcounter{table}{0}
\section{Supplementary information}
\subsection{Details on Numerical Calculations}
Density functional theory (DFT) calculations were performed using the Vienna Ab initio Simulation Package (VASP) \cite{kresse96_11169a, kresse96_15a}. The projector augmented-wave (PAW) method \cite{blochl94_17953a, kresse99_1758} was employed to describe the interactions between core and valence electrons, with valence configurations of Mo ($4s^2 4p^6 4d^5 5s^1$), S ($3s^2 3p^4$), and Se ($4s^2 4p^4$). The corresponding atomic radii are approximately 190, 88, and 103 pm, respectively \cite{clementi67_1300}. Exchange–correlation effects were treated using the Perdew–Burke–Ernzerhof (PBE) functional within the generalized gradient approximation (GGA), including van der Waals interactions via the D3 dispersion correction by Grimme (PBE+D3) \cite{perdew96_3865a, grimme10_154104}.  To improve the accuracy of band gaps, hybrid functional calculations were also performed using the Heyd–Scuseria–Ernzerhof functional (HSE06), which incorporates a screened fraction of exact exchange. 

For the GW and BSE calculations, we employed PAW pseudopotentials specifically optimized for many-body perturbation theory (GW PAW potentials), which include additional semicore states in the valence configuration. This ensures a more accurate description of the high-lying conduction bands and the screened Coulomb interaction, both of which are critical for obtaining reliable quasiparticle corrections and excitonic spectra. In contrast, standard PBE pseudopotentials were used for structural relaxations and ground-state electronic structure calculations.

A plane-wave energy cutoff of 450 eV was used for all calculations. The Brillouin zone was sampled using a $\Gamma$-centered Monkhorst–Pack mesh of $12 \times 12 \times 1$ \textbf{k}-points. A vacuum spacing greater than 20 \AA\ was included along the out-of-plane direction to prevent spurious interlayer interactions. All atomic positions were relaxed until the residual forces were below 0.01 eV Å$^{-1}$ and the total energy was converged to within $10^{-8}$ eV.

Ferroelectric polarization was computed using the  theory of polarization within the Berry phase formalism \cite{king-smith93_1651, resta94_899, xiao10_1959}, employing a dense $16 \times 16 \times 1$ \textbf{k}-point mesh. The macroscopic polarization $\mathbf{P}$ was evaluated as:

\begin{equation}
	\mathbf{P} = -\frac{e}{(2\pi)^3} \sum_n \int_{\text{BZ}} d\mathbf{k} \, \langle u_{n\mathbf{k}} | \nabla_{\mathbf{k}} | u_{n\mathbf{k}} \rangle,
	\label{eq:berry_phase_pol}
\end{equation}

where \( e \) is the elementary charge, \( u_{n\mathbf{k}} \) is the periodic part of the Bloch function for the \( n \)th occupied band, and the integral is performed over the Brillouin zone (BZ).

The Berry curvature $\Omega_z(\mathbf{k})$ was evaluated using the Kubo-like formula:

\begin{equation}
	\Omega_z(\mathbf{k}) = -\sum_{n} f_n \sum_{n' \ne n} \frac{2\, \mathrm{Im} \left[ \langle \psi_{n\mathbf{k}} | v_x | \psi_{n'\mathbf{k}} \rangle \langle \psi_{n'\mathbf{k}} | v_y | \psi_{n\mathbf{k}} \rangle \right]}{(E_{n'} - E_n)^2},
	\label{eq:berry_curvature}
\end{equation}

where $v_x$ and $v_y$ are the velocity operators in the $x$ and $y$ directions, $f_n$ is the Fermi–Dirac occupation factor, and $E_n$ are the Kohn–Sham eigenvalues.


Many-body perturbation theory was employed to calculate quasiparticle and optical excitation properties. Quasiparticle energies were obtained using single-shot G$_0$W$_0$  calculations by solving the equation \cite{hedin65_139A796, hybertsen1985_prl}:

\begin{equation}
	\varepsilon_i^{\text{QP}} = \varepsilon_i^{\text{KS}} + \langle \phi_i^{\text{KS}} | \Sigma(\varepsilon_i^{\text{QP}}) - V_{\text{XC}}^{\text{KS}} | \phi_i^{\text{KS}} \rangle,
\end{equation}

where \( \varepsilon_i^{\text{KS}} \) and \( \phi_i^{\text{KS}} \) are the Kohn–Sham eigenvalues and eigenfunctions, \( \Sigma \) is the nonlocal self-energy operator, and \( V_{\text{XC}}^{\text{KS}} \) is the exchange–correlation potential from the underlying DFT calculations.

Optical absorption spectra were calculated by solving the Bethe–Salpeter equation (BSE), which accounts for correlated electron–hole excitations \cite{hanke1980_prb, strinati1988_rnc}. The effective two-particle Hamiltonian is given by:

\begin{equation}
	H^{\text{BSE}} = H_{\text{diag}} + H_{\text{dir}} + 2H_{x},
\end{equation}

where \( H_{\text{diag}} \) corresponds to non-interacting quasiparticle transitions, \( H_{\text{dir}} \) accounts for the screened Coulomb attraction between the electron and hole (excitonic effects), and \( H_x \) is the bare exchange term, with the factor of two reflecting spin multiplicity in non-spin-polarized systems. The eigenvalues \( E^\lambda \) represent the excitation energies, and the corresponding eigenvectors \( A^\lambda_{vc\mathbf{k}} \) indicate the contributions from individual electron–hole transitions.

The oscillator strength for each excitation is given by:

\begin{equation}
	t_{\lambda} = \sum_{vc\mathbf{k}} A^{\lambda}_{vc\mathbf{k}} \frac{\langle v\mathbf{k} | \hat{\mathbf{p}} | c\mathbf{k} \rangle}{\varepsilon_{c\mathbf{k}} - \varepsilon_{v\mathbf{k}}},
\end{equation}

which contributes to the imaginary part of the macroscopic dielectric function:

\begin{equation}
	\mathrm{Im}\, \varepsilon_M(\omega) = \frac{8\pi^2}{\Omega} \sum_{\lambda} |t_{\lambda}|^2 \delta(\omega - E^{\lambda}),
\end{equation}

where \( \Omega \) is the unit cell volume.

To avoid image interactions in the non-periodic direction, a Coulomb truncation scheme was employed. All G$_0$W$_0$  and BSE calculations were performed using a $\Gamma$-centered \(12 \times 12 \times 1\) \textbf{k}-point grid. The number of unoccupied bands (\texttt{NBANDS}) was set to 32 for bilayer MoS$_2$ and MoS$_2$/MoSe$_2$, and increased to 64 for the trilayer 2MoS$_2$/MoSe$_2$ systems. All convergence parameters, including energy cutoffs and the number of bands, were carefully tested to ensure the reliability of the computed quasiparticle energies and excitonic spectra.

\subsection{Additional Computational Results}

\begin{figure*}[ht!]
	\centering
	\includegraphics[width=0.95\textwidth]{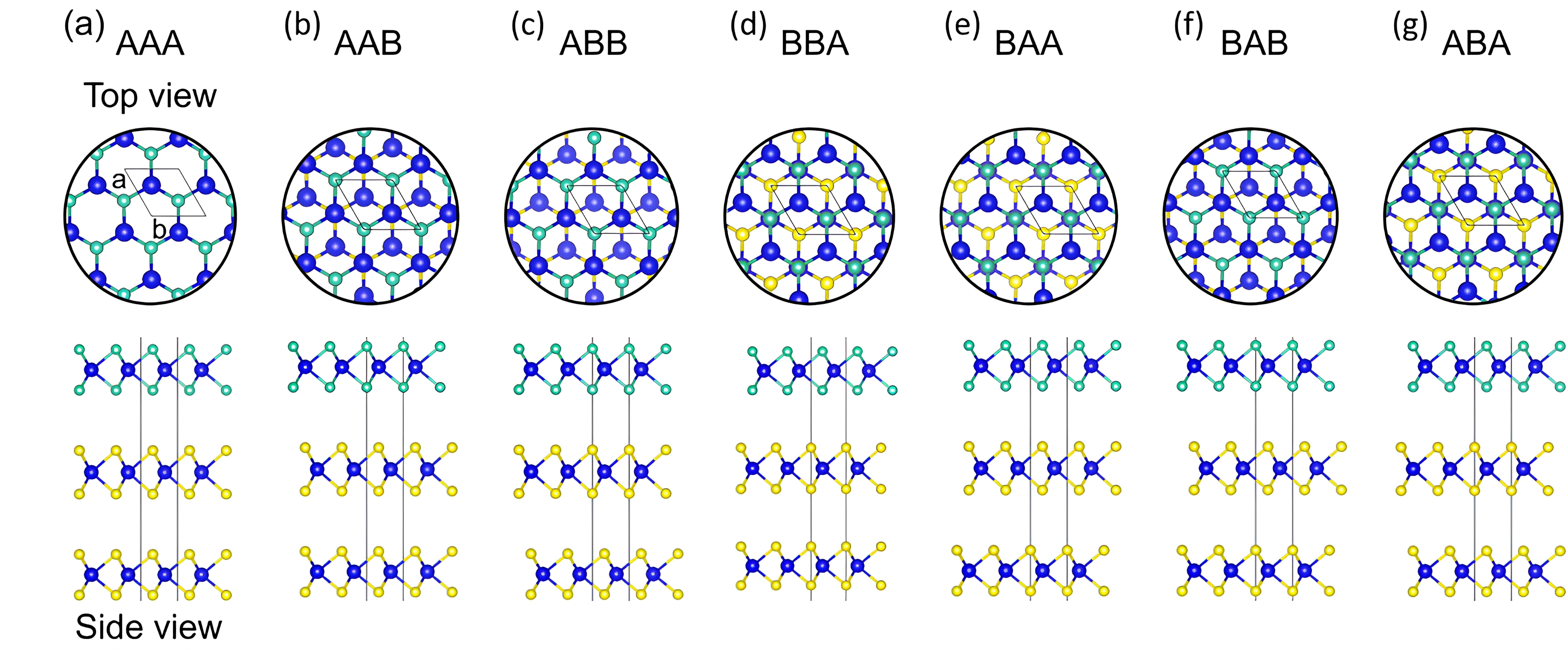}
	\caption{
		Top and side views of atomic structures for the 2L-MoS$_2$/MoSe$_2$ heterostructure in seven high-symmetry stacking configurations. 
		Panels (a)–(g) correspond to the AAA, AAB, ABB, BBA, BAA, BAB, and ABA stackings, respectively, each with a unique interlayer atomic registry. 
		These variations in stacking order affect the interlayer coupling and consequently the excitonic behavior of the heterostructure.
	}
	\label{fig:S1}
\end{figure*}

\begin{figure}[ht!]
	\centering
	\includegraphics[width=0.85\textwidth]{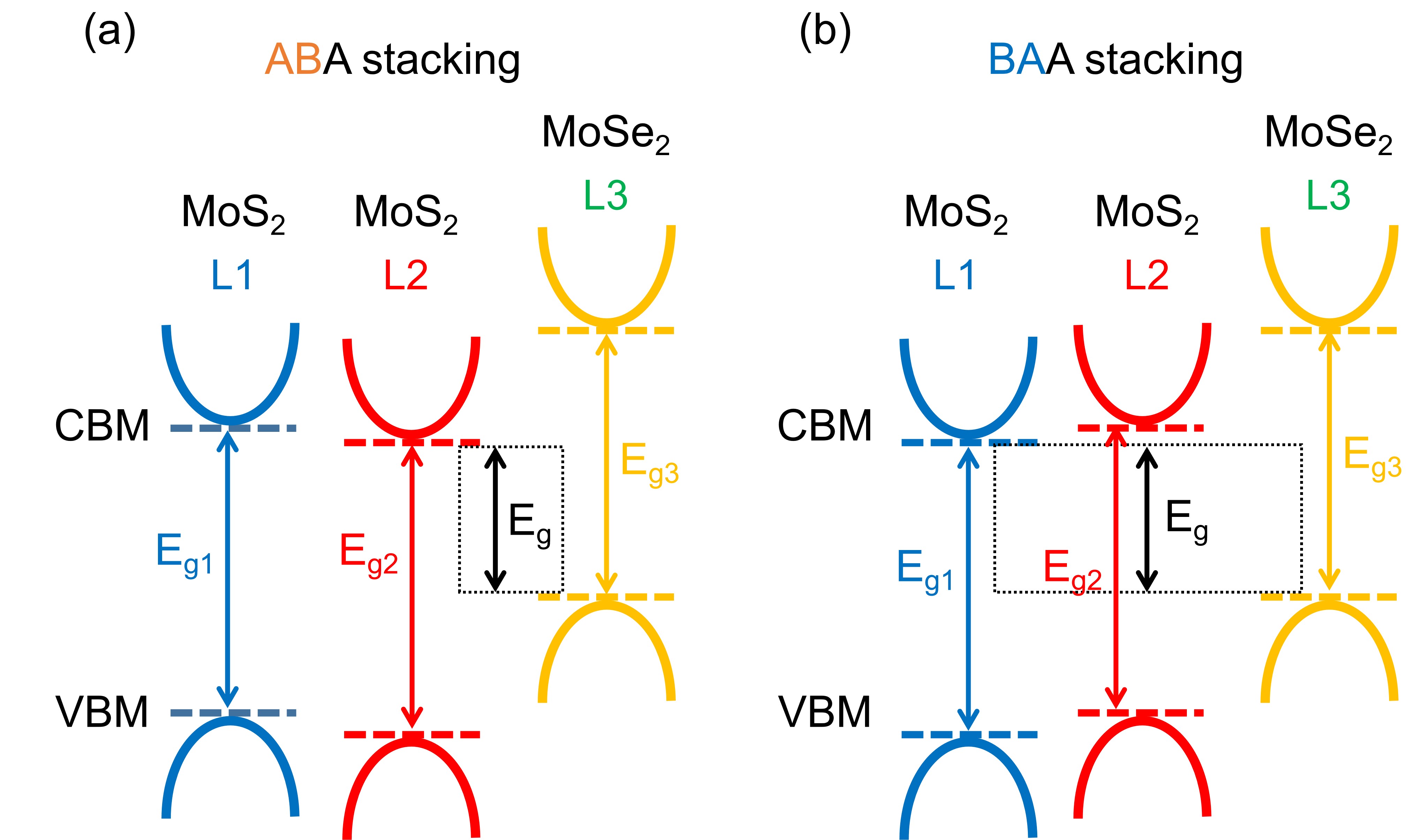}
	\caption{
		Band alignment of the 2L-MoS$_2$/MoSe$_2$ heterostructure in (a) ABA and (b) BAA stacking configurations. The system consists of two monolayers of MoS$_2$ (L1 and L2) and a monolayer of MoSe$_2$ (L3). Both stackings exhibit type-II band alignment, with the conduction band minimum (CBM) located on MoS$_2$ and the valence band maximum (VBM) on MoSe$_2$. In the ABA configuration, the band gap (E$_\mathrm{g}$) of the heterostructure is defined by the CBM of L2 MoS$_2$ and the VBM of L3 MoSe$_2$, while in the BAA configuration, E$_\mathrm{g}$ is defined by the CBM of L1 MoS$_2$ and the VBM of L3 MoSe$_2$. The individual band gaps of L1 (E$_\mathrm{g1}$), L2 (E$_\mathrm{g2}$), and L3 (E$_\mathrm{g3}$) are also indicated.
	}
	\label{fig:S2}
\end{figure}

\begin{table}[ht]
	\centering
	\caption{
		Calculated PL peak positions and first bright exciton energies with oscillator strengths (in parentheses)  for all stacking configurations of the 2L-MoS$_2$/MoSe$_2$ heterostructure.
	}
	\label{tab:PL}
	\begin{tabular}{l|c|c}
		\hline\hline
		{Stacking} & {PL Energy (eV)  } & {First Bright Exciton (eV) } \\
		\hline
		AAA & 1.362 (0.024) & 1.925 (13.23) \\
		AAB & 1.398 (0.003) & 1.893 (13.41) \\
		ABB & 1.391 (0.002) & 1.897 (13.38) \\
		BBA & 1.278 (0.069) & 1.887 (13.43) \\
		BAA & 1.356 (0.029) & 1.909 (13.68) \\
		BAB & 1.395 (0.002) & 1.894 (13.43) \\
		ABA & 1.287 (0.068) & 1.905 (13.79) \\
		\hline\hline
	\end{tabular}
\end{table}

\clearpage

\subsection{Observation of domain dependent ILX PL in an additional sample}

\begin{figure}
	\centering
	\includegraphics[width=\textwidth]{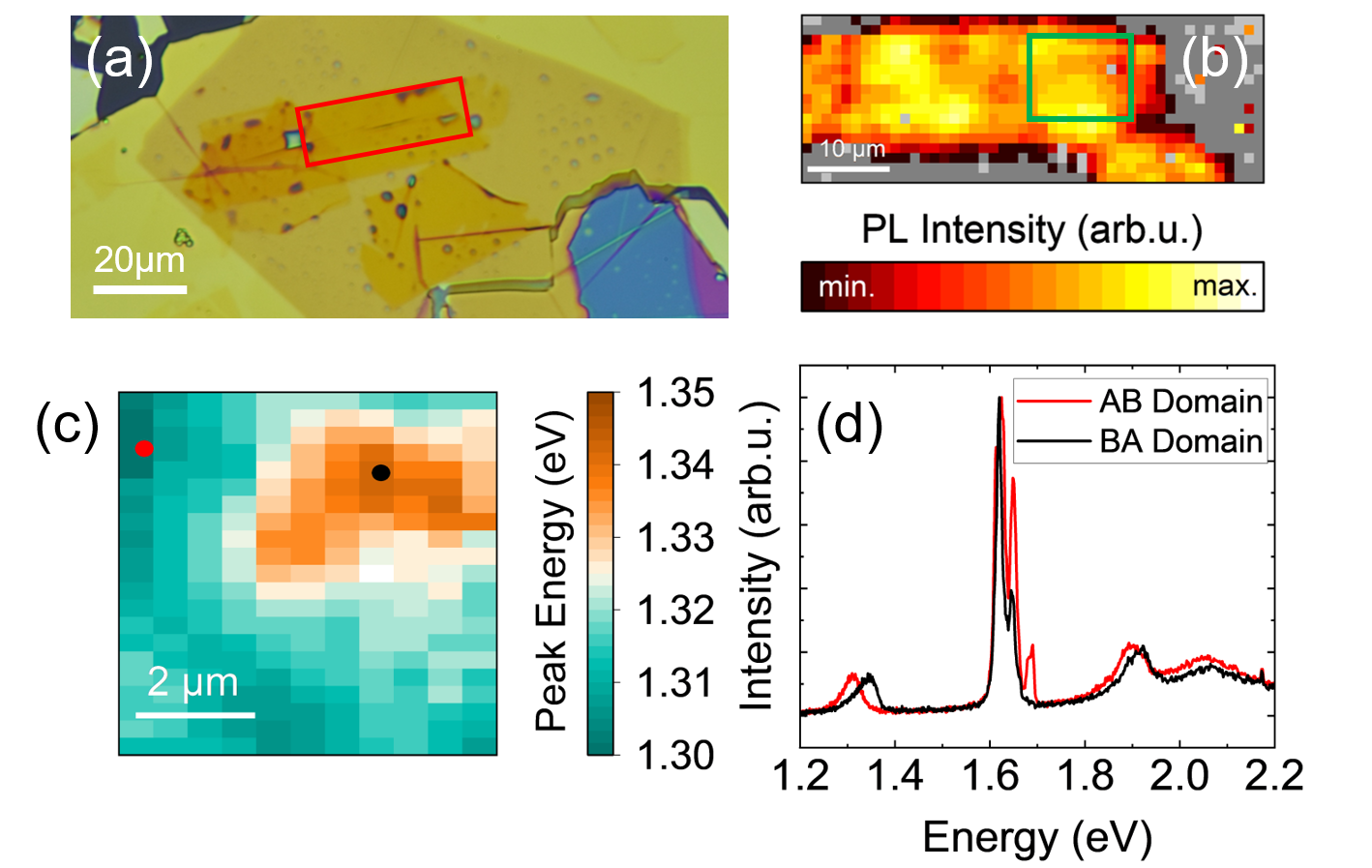}
	\captionsetup{width=\textwidth}
	\caption {(a) Optical image of an hBN-encapsulated 2L MoS$_{2}$/MoSe$_{2}$ heterostructure on a SiO$_{2}$ substrate. The red rectangle denotes the region of the PL intensity of the ILX map shown in (b). The green square in (b) marks the PL peak energy map of the ILX in (c). In (d) two exemplary spectra from the map of (c) are shown. The energy difference of the ILX between the domains approximates to 40~meV.}
	\label{S_Panel1}
\end{figure}

In Fig.\ref{S_Panel1}a another hBN-encapsulated 2L MoS$_{2}$/MoSe$_{2}$ heterostructure is shown. It has the same stacking configuration as the one shown in fig. 2 of the main text. Again, an ILX resonance can be seen in the whole heterostructure. The ILX shows also an energy shift of about 40~meV in the ILX energy depending on the location of the sample, while the other features shift only minimally (see fig. \ref{S_Panel1}d). The different energies correspond to the different ferroelectric domains in the 3R-MoS$_{2}$ bilayer.  

\clearpage

\subsection{AFM measurements}
\begin{figure}
	\centering
	\includegraphics[width=\textwidth]{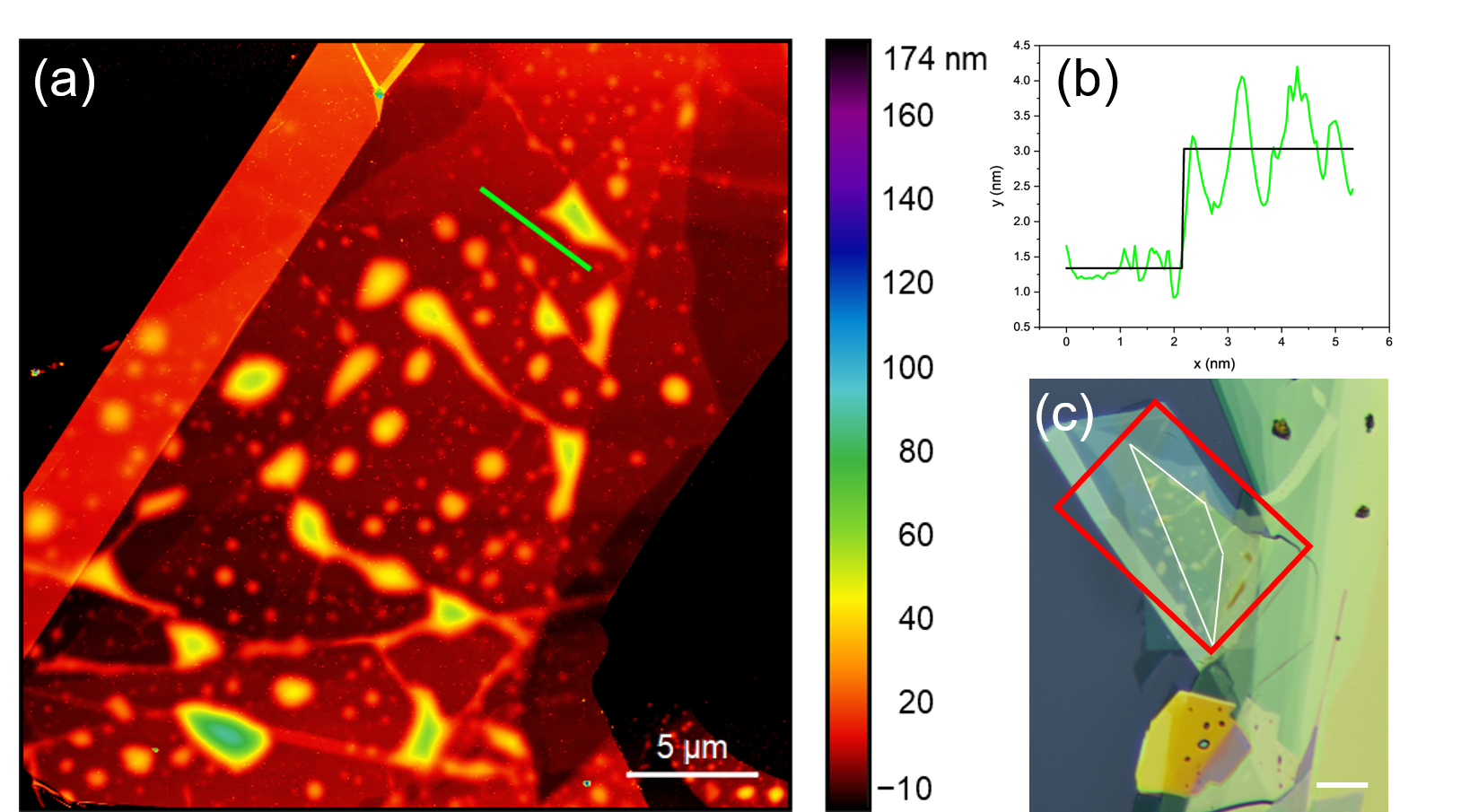}
	\captionsetup{width=\textwidth}
	\caption {(a) AFM scan on sample from Fig.2. The green line denotes the profile shown in (b). The step height fitted approximates to 1.4~nm. (c) Optical image of the sample with the red indicator showing the size of the AFM scan (a). The white lines corresponds to roughly 5µm.}
	\label{S_Panel2}
	
\end{figure}

In \ref{S_Panel2} the AFM topography of the sample analyzed in Fig. 2 is shown. The green line in Fig~\ref{S_Panel2}a and \ref{S_Panel2}b depicts the profile line of the 2L MoS$_{2}$ compared from the MoSe$_{2}$ monolayer to the MoS$_{2}$/MoSe$_{2}$ heterostructure. The fit estimates roughly to 1.4~nm, corroborating the bilayer thickness of the MoS$_{2}$.
\clearpage
\section{Redshift of the ILX in different heterostructures}

\begin{table}[h]
	\centering
	\caption{Energies of different 3R-MoS$_{2}$/MoSe$_{2}$ heterostructures with varying MoS$_{2}$ layer counts.}
	\begin{tabular}{lcc}
		\toprule
		Heterostructure & $E^{ILX}_{exp}$ (eV) & $E^{ILX}_{theory}$ (eV) \\
		\midrule
		1L-MoS$_{2}$/MoSe$_{2}$ & 1.42 & 1.44 \\
		2L-MoS$_{2}$/MoSe$_{2}$ & 1.34 & 1.36 \\
		3L-MoS$_{2}$/MoSe$_{2}$ &  1.24 & -- \\
		4L-MoS$_{2}$/MoSe$_{2}$ & 1.16 &  -- \\
		\bottomrule
	\end{tabular}
\end{table}
\clearpage

\begin{figure}
	\centering
	\includegraphics[width=0.5\linewidth]{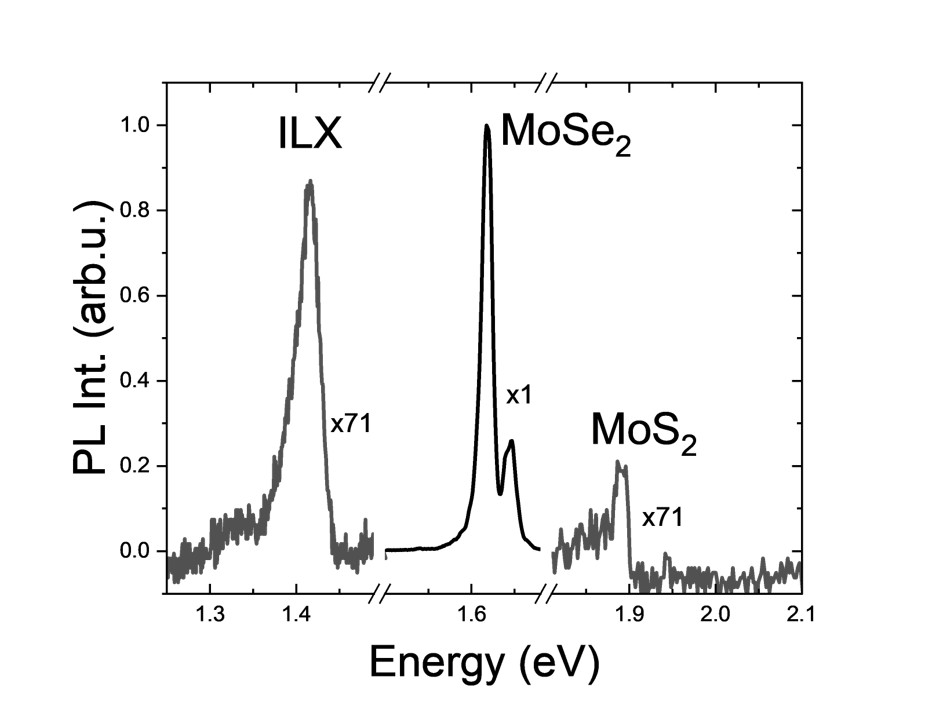}
	\captionsetup{width=0.5\linewidth}
	\caption {Exemplary spectrum from a 1L-MoS$_{2}$/MoSe$_{2}$ heterostructure at 4~K.}
	\label{S_Panel3}
\end{figure}
\clearpage
\subsection{Second Harmonic Generation}
\begin{figure}
	\centering
	\includegraphics[width=1\linewidth]{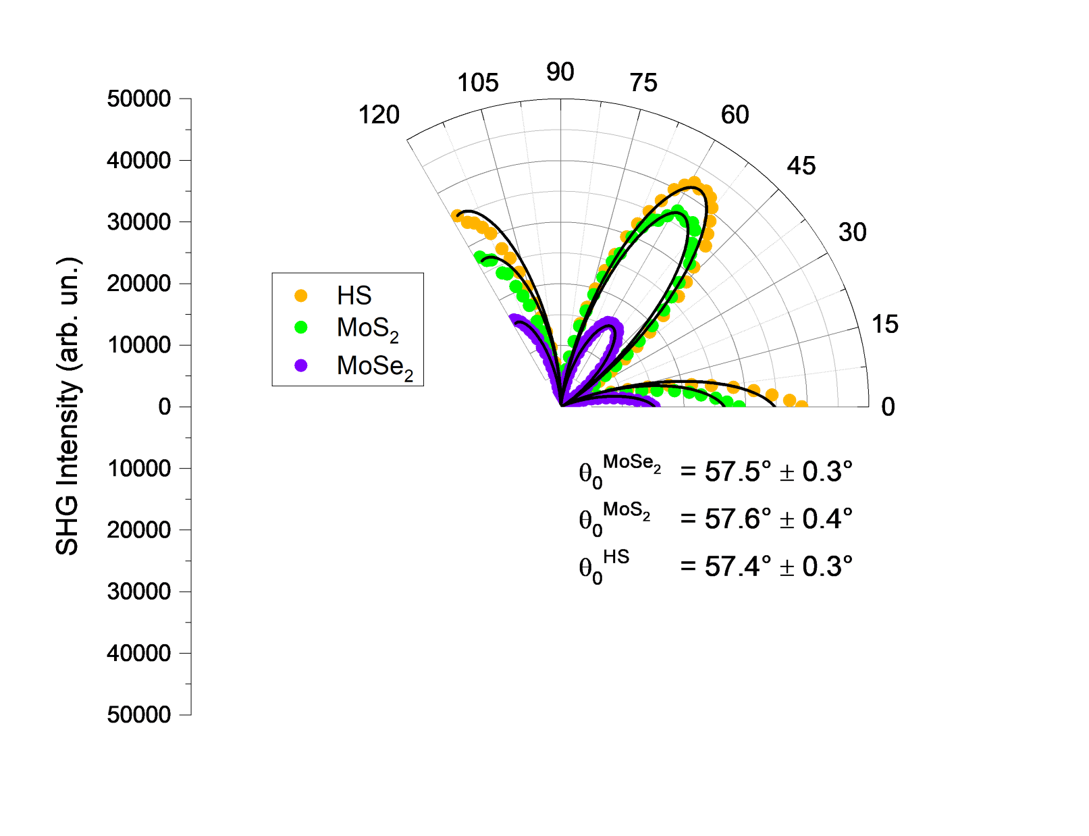}
	\captionsetup{width=1\linewidth}
	\caption {SHG intensity of the sample from Fig.~2.}
	\label{S_Panel4}
\end{figure}

In \ref{S_Panel4} the SHG intensity of the sample analyzed in Fig. 2 is shown. The SHG confirms the good alignment between MoS$_{2}$ and MoSe$_{2}$ with $\Delta \theta \approx$ 0.1~$^\circ$. Also by comparison of the SHG intensity of the individual layers and the heterostructure, a R-stacking can be deduced. 

\end{document}